\documentclass[10pt,twocolumn,letterpaper]{article}

 \usepackage{cvpr}              % To produce the CAMERA-READY version
\usepackage{makecell}
\usepackage{amsmath}
\usepackage{graphicx}
\usepackage{caption}
\usepackage{subcaption}
\usepackage{comment}
\usepackage{upgreek}
\usepackage{multirow}
\usepackage{mathtools}
\usepackage[utf8]{inputenc} % allow utf-8 input
\usepackage[T1]{fontenc}    % use 8-bit T1 fonts
\usepackage{url}            % simple URL typesetting
\usepackage{booktabs}       % professional-quality tables
\usepackage{amsfonts}       % blackboard math symbols
\usepackage{nicefrac}       % compact symbols for 1/2, etc.
\usepackage{microtype}      % microtypography
\usepackage{xcolor}         % colors
\usepackage{adjustbox}
\usepackage[margin=1in]{geometry}
\usepackage[symbol]{footmisc}
\definecolor{cvprblue}{rgb}{0.21,0.49,0.74}
\usepackage[pagebackref,breaklinks,colorlinks,allcolors=cvprblue]{hyperref}

\def\paperID{17} % *** Enter the Paper ID here
\def\confName{CVPR}
\def\confYear{2025}

\title{Cultural Awareness in Vision-Language Models: A Cross-Country Exploration}

\newcommand{\intel}{$^{\heartsuit}$}
\newcommand{\tw}{$^{\diamondsuit}$}

\author{
  Avinash Madasu\intel \qquad
  Vasudev Lal\intel \qquad
  \stepcounter{footnote} Phillip Howard\tw\thanks{Work completed while at Intel Labs.} \\
  \intel Intel Labs \qquad
  \tw Thoughtworks \\
  \qquad {\tt\small \{avinash.madasu, vasudev.lal}\tt\small\}@intel.com \qquad phillip.howard@thoughtworks.com
}

\begin{document}
\maketitle

\begin{abstract}
Vision-Language Models (VLMs) are increasingly deployed in diverse cultural contexts, yet their internal biases remain poorly understood. In this work, we propose a novel framework to systematically evaluate how VLMs encode cultural differences and biases related to race, gender, and physical traits across countries. We introduce three retrieval-based tasks: (1) Race to Country retrieval, which examines the association between individuals from specific racial groups (East Asian, White, Middle Eastern, Latino, South Asian, and Black) and different countries; (2) Personal Traits to Country retrieval, where images are paired with trait-based prompts (e.g., Smart, Honest, Criminal, Violent) to investigate potential stereotypical associations; and (3) Physical Characteristics to Country retrieval, focusing on visual attributes like skinny, young, obese, and old to explore how physical appearances are culturally linked to nations. Our findings reveal persistent biases in VLMs, highlighting how visual representations may inadvertently reinforce societal stereotypes.
\end{abstract}
  
\section{Introduction}
\label{sec:intro}

Vision-Language Models (VLMs) ~\cite{radford2021learning, yang2023alip, fan2023improving, cherti2023reproducible, li2023blip, zhai2023sigmoid, singh2022flava}, pre-trained on large-scale image-text datasets, have achieved state-of-the-art results on a variety of tasks, including image retrieval, captioning, and visual question answering (VQA). Despite these advancements, recent studies ~\cite{zhao2021understanding, garcia2023uncurated, birhane2021multimodal, howard2024socialcounterfactuals, zhou2022vlstereoset, hall2023vision, janghorbani2023multi} have highlighted the presence of gender, racial, and intersectional social biases in these models. As VLMs are increasingly deployed in socially and culturally sensitive contexts across the globe, ensuring that they capture a broader and more inclusive understanding of diverse cultures has become imperative. However, prior work ~\cite{shankarno, nayak2024benchmarking, liu2021visually, pouget2024no, nwatu2023bridging, luccioni2023stable} has demonstrated that VLMs exhibit a strong bias toward Western-centric concepts in tasks such as text-to-image generation, image captioning, and object detection. In response, recent efforts ~\cite{bhatia2024local, ananthramsee, nayak2024benchmarking} have introduced novel evaluation benchmarks to assess the multicultural competence of VLMs. Nonetheless, these benchmarks are often limited to evaluating between western and non-western countries. However, this binary choice of evaluation is reductive in nature often ignoring multiple cultures.

In this work, we propose a comprehensive evaluation framework to assess the cultural understanding of Vision-Language Models (VLMs). Specifically, we focus on six major racial and ethnic groups\footnote{We use the terms race and ethnicity loosely in this work}: White, Black, South Asian, East Asian, Middle Eastern, and Latino, selecting the top ten countries with significant populations for each group. Using the FairFace ~\cite{karkkainen2021fairface} and Social Counterfactuals ~\cite{howard2024socialcounterfactuals} datasets, we evaluate the distribution of countries retrieved by VLMs given an image of a person from a specific racial group. This allows us to investigate whether VLMs can accurately associate visual appearances with culturally appropriate geographic contexts. Additionally, we extend this analysis by pairing images with text prompts describing personal traits (e.g., honest, lazy, criminal, etc.)  and physical characteristics (e.g., skinny, obese, old, etc.) attributed to different countries, and measure the proportion of countries associated with each racial group. This enables us to examine potential stereotypes or cultural biases embedded within VLMs when associating personal and physical traits with race and nationality. 

\begin{table}[]
    \scriptsize
    \centering
    \begin{tabular}{cc}
    \toprule
    \textbf{Race /Ethnicity} & \textbf{Countries} \\
    \toprule
      East asian   &  \makecell{China, Indonesia, Japan, \\ Philippines, Vietnam, Thailand, \\ Myanmar, South Korea, Malaysia, \\ North Korea} \\
      \toprule
       White (Caucasian)  & \makecell{United States, Russia, Brazil, \\ Germany, France, Italy,\\ United Kingdom, Spain, \\ Ukraine, Argentina} \\
       \toprule
       Middle Eastern & \makecell{Egypt, Iran, Turkey, Iraq, \\ Saudi Arabia, Yemen, Syria, \\ Jordan, United Arab Emirates, \\ Israel} \\
       \toprule
       Latino & \makecell{Brazil, Colombia, Argentina, \\ Peru, Venezuela, \\ Chile, Ecuador, \\Bolivia, Paraguay, Uruguay} \\
       \toprule
       South Asian & \makecell{India, Pakistan, \\ Bangladesh, Nepal, Sri Lanka, \\Afghanistan, Bhutan, Maldives} \\
       \toprule
       Black & \makecell{Nigeria, Ethiopia, \\ Democratic Republic of Congo, \\ United States, Brazil, \\ South Africa, Tanzania, \\ Kenya, Uganda, Ghana} \\
       \toprule
    \end{tabular}
    \caption{List of countries covered in the retrieval task for each race or ethnicity.}
    \label{tab:race_countries}
\end{table}

\begin{table}
\footnotesize
    \begin{tabular}{cc}
    \toprule
       \textbf{Task}  &  \textbf{Prompt} \\
       \toprule
      Race to Country   & A person from China. \\
      Physical Characteristics to Country & A skinny person from Peru. \\
      Physical Traits to Country retrieval & A talented person from Kenya. \\
      \toprule
    \end{tabular}
    \caption{Example prompts used in different cultural tasks.}
    \label{tab:task_prompt}
\end{table}

\section{Methodology}
\subsection{Task Definition}
 We propose a new task, namely \textbf{Retrieval across Countries}, aimed at evaluating the geographical and cultural understanding of Vision-Language Models (VLMs). Specifically, given an input image $I$ of a person and a set of textual prompts describing various countries $T = {T_1, T_2, T_3, ..., T_n}$, the goal is to identify the country that the VLM most closely associates with the input image. We then measure the proportion of total images associated with a particular country, denoted as:
\[
\text{R@C} = \frac{\sum_{i=1}^{N} \mathbb{1}\left[ \arg\max_{j \in \mathcal{C}} \text{Sim}(I_i, T_j) = C \right]}{N}
\]
where N is the total number of images, C is the set of all country text embeddings, $I_{j}$ is image embedding for the $i^{th}$ image, $Sim(I_{i}, T_{j})$ is the similarity score between image $I_{i}$ and $T_{j}$ and \(\mathbb{1}[\cdot]\) is an indicator function.
\subsection{Tasks} We focus on three distinct tasks to evaluate the cultural understanding and potential biases exhibited by Vision-Language Models (VLMs). The  countries and prompts used for each task are are shown in Tables ~\ref{tab:race_countries} and ~\ref{tab:task_prompt} respectively. \\
\textbf{Race to Country retrieval:} This task measures how often VLMs associate images of individuals from specific racial groups (East Asian, White, Middle Eastern, Latino, South Asian, and Black) with different countries. It uncovers potential racial biases and stereotypes in the models' learned representations. \\
\textbf{Personal Traits to Country retrieval:} This task pairs images with prompts describing traits like Smart, Honest, Criminal, and Violent, and observes which countries are retrieved. It explores whether VLMs exhibit biases by linking positive or negative personal traits to specific countries or racial groups. \\
\textbf{Physical Characteristics to Country retrieval:} In this task, images based on physical attributes like skinny, young, obese, and old are used to analyze how VLMs associate physical appearance with countries. It helps reveal culturally-ingrained visual biases linked to body types and age. \\
\subsection{Selection of Countries} For each racial/ethnic group, we select the top-10 most populated countries where these groups are predominantly found. Table~\ref{tab:race_countries} lists the countries considered in this study. This ensures a globally representative evaluation, focusing on regions with significant demographic presence for each racial/ethnic group, and providing a balanced basis for analyzing cultural bias in VLMs.
\subsection{Datasets and Models}
\textbf{Datasets:} FairFace ~\cite{karkkainen2021fairface} and SocialCounterFactuals ~\cite{howard2024socialcounterfactuals}. \\
\textbf{Models:} ALIP ~\cite{yang2023alip}, LACLIP ~\cite{fan2023improving}, OpenCLIP ~\cite{cherti2023reproducible} and BLIP-2 ~\cite{li2023blip}.

\begin{figure*}[htbp]
    \centering

    \begin{subfigure}[t]{0.49\textwidth}
        \includegraphics[width=\linewidth]{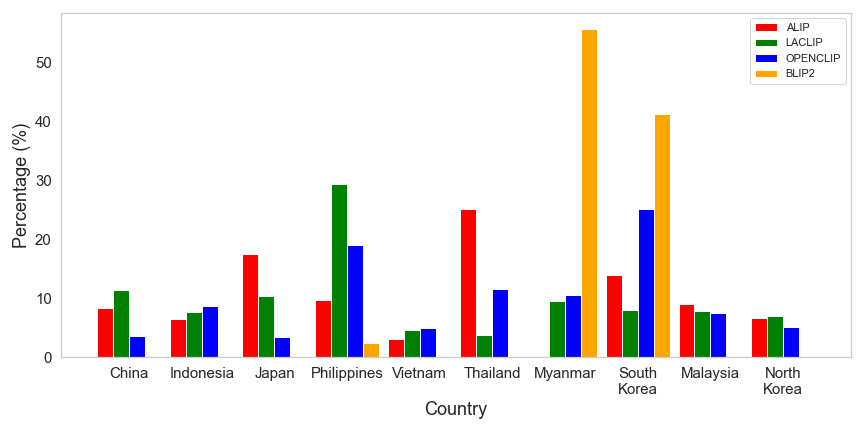}
        \caption{East Asian}
    \end{subfigure}
    \hfill
    \begin{subfigure}[t]{0.49\textwidth}
        \includegraphics[width=\linewidth]{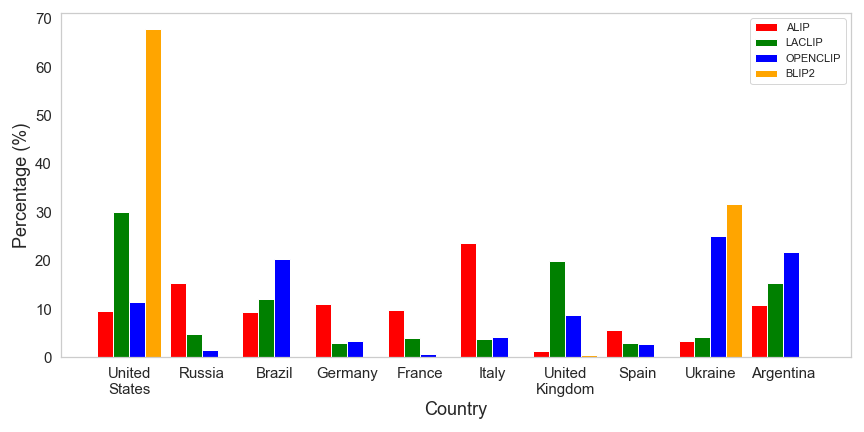}
        \caption{White}
    \end{subfigure}

    \vspace{0.4cm}

    \begin{subfigure}[t]{0.49\textwidth}
        \includegraphics[width=\linewidth]{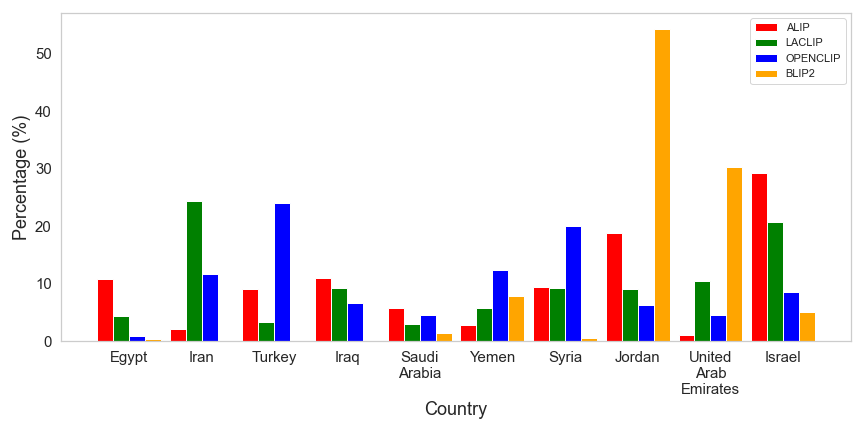}
        \caption{Middle Eastern}
    \end{subfigure}
    \hfill
    \begin{subfigure}[t]{0.49\textwidth}
        \includegraphics[width=\linewidth]{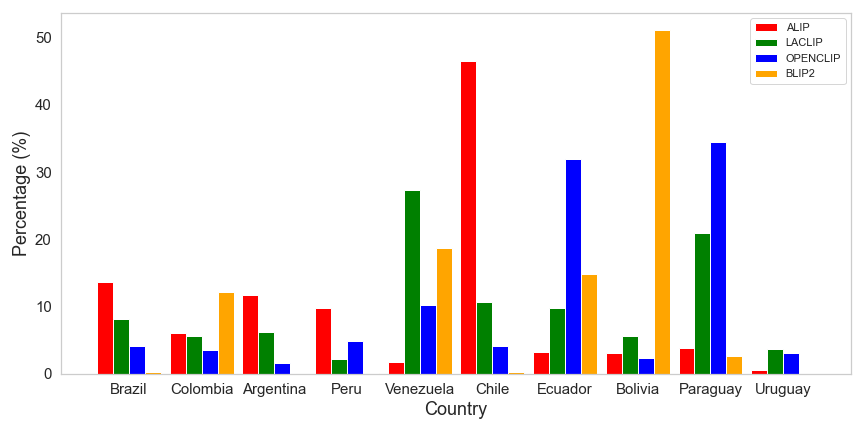}
        \caption{Latino}
    \end{subfigure}

    \vspace{0.4cm}

    \begin{subfigure}[t]{0.49\textwidth}
        \includegraphics[width=\linewidth]{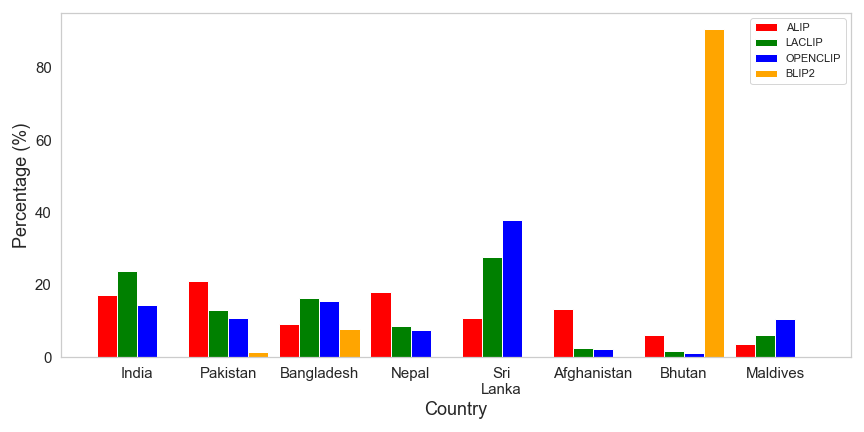}
        \caption{South Asian}
    \end{subfigure}
    \hfill
    \begin{subfigure}[t]{0.49\textwidth}
        \includegraphics[width=\linewidth]{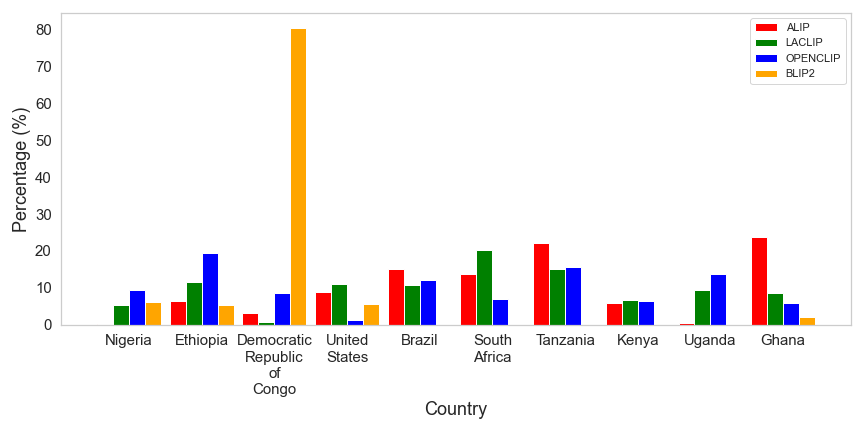}
        \caption{Black}
    \end{subfigure}

    \caption{\textbf{Performance comparison of different VLMs on various races and retrieved countries as a percentage of total images on FairFace dataset.}}
    \label{fig:race_country_fairface}
\end{figure*}

\begin{table*}[t]
\centering
\tiny
\setlength{\tabcolsep}{1.8pt} % Reduces column separation
\renewcommand{\arraystretch}{1.3} % Adjust row spacing slightly if needed
\begin{tabular}{l|*{6}{c}|*{6}{c}}
\toprule
& \multicolumn{6}{c|}{ALIP} & \multicolumn{6}{c}{LACLIP} \\
\cline{2-13}
Trait & \makecell{East\\Asian} & \makecell{South\\Asian} & White & \makecell{Middle\\Eastern} & Latino & Black 
      & \makecell{East\\Asian} & \makecell{South\\Asian} & White & \makecell{Middle\\Eastern} & Latino & Black \\
\midrule
Smart & Philippines (45.05)  & Nepal (29.22)  & Russia (44.94)  & Egypt (33.21)  & Chile (67.68)  & Ethiopia (35.97)  & Philippines (29.49)  & SriLanka (26.73)  & Ukraine (22.77)  & Iran (21.18)  & Venezuela (30.93)  & South Africa (21.92) 
\\
Honest & Philippines (40.4)  & SriLanka (31.81)  & Argentina (33.88)  & Jordan (34.16)  & Argentina (33.01)  & South Africa (46.79)  & Philippines (21.05)  & SriLanka (27.46)  & Ukraine (26.74)  & Iran (25.52)  & Paraguay (22.73)  & South Africa (19.52) 
\\
Successful & North Korea (19.34)  & Nepal (31.37)  & Spain (27.71)  & Egypt (43.73)  & Chile (46.34)  & South Africa (27.65)  & Philippines (29.88)  & Bangladesh (23.52)  & Ukraine (21.38)  & Iran (20.12)  & Venezuela (23.32)  & Ethiopia (19.73) 
\\
Talented & Philippines (35.49)  & SriLanka (43.87)  & Germany (25.87)  & Jordan (22.56)  & Brazil (39.08)  & Brazil (28.87)  & Philippines (30.48)  & SriLanka (24.14)  & USA (19.49)  & Israel (19.08)  & Ecuador (22.65)  & Tanzania (20.51) 
\\
Creative & Japan (31.24)  & SriLanka (40.06)  & France (29.75)  & Israel (30.99)  & Chile (45.79)  & Ethiopia (26.27)  & Philippines (34.48)  & SriLanka (29.29)  & Ukraine (28.75)  & Syria (23.24)  & Venezuela (35.29)  & Ethiopia (25.15) 
\\
Hardworking & China (16.9)  & Bhutan (23.53)  & Spain (24.92)  & SaudiArabia (40.92)  & Uruguay (27.65)  & Tanzania (28.64)  & Philippines (32.58)  & SriLanka (26.85)  & Ukraine (28.02)  & Iran (30.73)  & Venezuela (27.69)  & South Africa (24.26) 
\\
\bottomrule
Criminal & China (24.99)  & SriLanka (50.05)  & Germany (34.0)  & Israel (23.22)  & Chile (50.94)  & SouthAfrica (32.24)  & Philippines (21.09)  & India (29.15)  & Ukraine (17.6)  & Iran (23.14)  & Ecuador (16.63)  & Ethiopia (18.86) 
\\
Lazy & Japan (36.66)  & India (34.81)  & Germany (53.56)  & Israel (51.26)  & Chile (37.23)  & Ethiopia (49.49)  & Philippines (25.09)  & SriLanka (29.09)  & Ukraine (47.73)  & Iran (20.67)  & Venezuela (27.23)  & Ethiopia (23.53) 
\\
Dangerous & Japan (30.79)  & SriLanka (39.07)  & France (26.65)  & Jordan (33.5)  & Chile (37.4)  & SouthAfrica (37.98)  & Philippines (23.6)  & Bangladesh (23.04)  & Ukraine (22.13)  & Israel (23.6)  & Venezuela (31.08)  & SouthAfrica (27.49) 
\\
Poor & Philippines (29.72)  & Nepal (30.03)  & France (45.57)  & Jordan (36.28)  & Peru (30.3)  & SouthAfrica (58.94)  & Indonesia (14.92)  & SriLanka (23.35)  & USA (20.36)  & Iran (28.34)  & Venezuela (40.09)  & SouthAfrica (32.8) 
\\
Violent & China (25.9)  & SriLanka (40.97)  & Argentina (61.52)  & Israel (37.41)  & Argentina (39.5)  & Kenya (23.73)  & Philippines (18.14)  & India (25.38)  & USA (17.9)  & Iran (24.86)  & Venezuela (15.94)  & Ethiopia (20.74) 
\\
Illiterate & Japan (26.15)  & Pakistan (21.21)  & Ukraine (34.66)  & Israel (23.38)  & Paraguay (28.44)  & SouthAfrica (24.26)  & Philippines (21.77)  & India (22.7)  & Ukraine (30.2)  & Iran (22.25)  & Venezuela (32.24)  & SouthAfrica (22.82) 
\\
\bottomrule
\end{tabular}
\caption{\textbf{Comparison of personal traits across demographics for ALIP and LACLIP models on FairFace dataset.}}
\label{tab:trait_comparison_alip_laclip}
\end{table*}

\begin{table*}[t]
\centering
\tiny
\setlength{\tabcolsep}{1.8pt} % Reduces column separation
\renewcommand{\arraystretch}{1.3} % Adjust row spacing slightly if needed
\begin{tabular}{l|*{6}{c}|*{6}{c}}
\toprule
& \multicolumn{6}{c|}{OpenCLIP} & \multicolumn{6}{c}{BLIP-2} \\
\cline{2-13}
Trait & \makecell{East\\Asian} & \makecell{South\\Asian} & White & \makecell{Middle\\Eastern} & Latino & Black 
      & \makecell{East\\Asian} & \makecell{South\\Asian} & White & \makecell{Middle\\Eastern} & Latino & Black \\
\midrule
Smart & Myanmar (20.96)  & SriLanka (25.36)  & Brazil (25.19)  & Syria (28.08)  & Ecuador (31.45)  & Ethiopia (29.42)  & SouthKorea (61.75)  & Bhutan (80.83)  & Ukraine (54.03)  & Jordan (85.63)  & Colombia (63.17)  & Ethiopia (49.11) 
\\
Honest & Myanmar (27.53)  & Bangladesh (30.77)  & Argentina (30.91)  & Syria (30.77)  & Ecuador (57.65)  & Ethiopia (31.17)  & Myanmar (31.88)  & Bhutan (82.58)  & Ukraine (39.81)  & Jordan (98.59)  & Colombia (76.05)  & Ethiopia (71.45) 
\\
Successful & Myanmar (21.53)  & Bangladesh (25.25)  & Argentina (31.82)  & Syria (35.76)  & Ecuador (29.38)  & Ethiopia (24.05)  & Japan (48.71)  & Bhutan (84.9)  & Italy (30.59)  & Jordan (82.78)  & Bolivia (59.27)  & Ethiopia (61.65) 
\\
Talented & Philippines (18.3)  & SriLanka (26.07)  & Argentina (25.28)  & Syria (24.74)  & Paraguay (29.99)  & Ethiopia (23.0)  & Japan (76.79)  & SriLanka (46.65)  & UK (52.33)  & Jordan (97.71)  & Colombia (85.52)  & Ethiopia (74.85) 
\\
Creative & Philippines (21.15)  & SriLanka (31.46)  & Argentina (28.95)  & Iraq (23.1)  & Ecuador (35.36)  & Ethiopia (26.79)  & Myanmar (52.96)  & Bhutan (51.21)  & Germany (84.26)  & Jordan (98.4)  & Colombia (85.97)  & DRC (62.67) 
\\
Hardworking & Philippines (22.64)  & SriLanka (30.3)  & Ukraine (20.72)  & Syria (28.58)  & Ecuador (23.86)  & Brazil (24.12)  & Philippines (72.86)  & Bhutan (83.26)  & Ukraine (65.05)  & UAE (40.51)  & Colombia (53.41)  & DRC (91.03) 
\\
\bottomrule
Criminal & Philippines (16.41)  & SriLanka (31.21)  & Brazil (25.06)  & Turkey (23.78)  & Ecuador (24.0)  & Brazil (21.74)  & Philippines (94.79)  & Bhutan (85.59)  & USA (54.04)  & UAE (81.35)  & Ecuador (67.34)  & DRC (86.18) 
\\
Lazy & Philippines (17.82)  & India (36.23)  & Ukraine (21.27)  & Syria (25.95)  & Paraguay (26.83)  & Ethiopia (26.35)  & SouthKorea (57.65)  & Bhutan (98.86)  & Ukraine (36.21)  & Jordan (97.93)  & Colombia (55.99)  & Ethiopia (71.66) 
\\
Dangerous & SouthKorea (25.48)  & India (25.64)  & Argentina (23.38)  & Syria (22.27)  & Ecuador (32.25)  & Ethiopia (24.12)  & SouthKorea (53.54)  & Bhutan (90.72)  & Ukraine (59.24)  & Jordan (98.37)  & Colombia (98.43)  & Ethiopia (56.81) 
\\
Poor & Thailand (17.6)  & SriLanka (30.01)  & Argentina (20.86)  & Israel (22.79)  & Ecuador (30.05)  & Brazil (24.66)  & SouthKorea (50.91)  & Bhutan (85.33)  & Ukraine (51.3)  & Jordan (94.9)  & Colombia (64.49)  & DRC (88.44) 
\\
Violent & Philippines (15.86)  & India (30.04)  & Argentina (24.38)  & Iran (18.92)  & Ecuador (30.38)  & Ethiopia (20.72)  & Philippines (88.2)  & Bhutan (62.38)  & Ukraine (56.99)  & UAE (95.8)  & Colombia (95.8)  & DRC (59.36) 
\\
Illiterate & SouthKorea (18.36)  & SriLanka (32.67)  & Ukraine (25.29)  & Israel (29.16)  & Venezuela (21.05)  & Brazil (29.89)  & SouthKorea (85.42)  & Bhutan (71.69)  & Ukraine (83.46)  & SaudiArabia (49.15)  & Colombia (58.8)  & DRC (80.27) 
\\
\bottomrule
\end{tabular}
\caption{\textbf{Comparison of personal traits across demographics for OpenCLIP and BLIP-2 models on FairFace dataset.}}
\label{tab:trait_comparison_openclip_blip2}
\end{table*}

\section{Results}
\label{sec:results}
\subsection{Race to Country retrieval.}
To investigate implicit regional biases in vision-language models (VLMs), we conducted a pair of country retrieval analyses conditioned on racial appearance, using two complementary datasets: FairFace and SocialCounterfactuals. Figure ~\ref{fig:race_country_fairface} shows the results on FairFace dataset\footnote{Due to space limitation, results for socialcounterfactuals is in appendix}. In both cases, images of individuals from six racial categories—East Asian, White, Middle Eastern, Latino, South Asian, and Black—were paired with neutral country prompts, and the proportion of times each country was retrieved was measured. We evaluated four recent VLMs: ALIP, BLIP-2, LACLIP, and OpenCLIP.

In the FairFace setting, the results revealed distinct patterns of regional association that reflected both expected stereotypes and surprising biases. For example, BLIP-2 strongly associated White individuals with the United States (67\%) and Black individuals with the Democratic Republic of Congo (80\%), while OpenCLIP favored Ethiopia and Brazil for Black faces. LACLIP demonstrated more balanced retrieval across countries but still exhibited localized preferences, such as Iran for Middle Eastern and Venezuela for Latino identities. These findings highlighted the presence of learned socio-cultural and geographic biases within the latent representations of VLMs. We observe similar set of results on socialcounterfactuals dataset. Taken together, these analyses reveal that vision-language models encode stereotypical geographic associations. 

\subsection{Personal Traits to Country Retrieval.}
Next, we assess the cultural understanding of personal traits across racial groups using FairFace dataset. In an image-text matching setup, we prompted models with \textit{"A [trait] person from [country]"}, measuring the maximum retrieval probability across different races (East Asian, South Asian, White, Middle Eastern, Latino, Black) for both positive and negative personal traits. Tables ~\ref{tab:trait_comparison_alip_laclip} and ~\ref{tab:trait_comparison_openclip_blip2} show the results of this setup.

\subsubsection{Positive Traits}
\textbf{Regional Consistency}: ALIP and LACLIP consistently retrieved \textit{Philippines}, \textit{Sri Lanka}, and \textit{Nepal} for positive traits associated with East Asians and South Asians, respectively. Similarly, positive associations for Black individuals centered around \textit{Ethiopia}, \textit{South Africa}, and \textit{Brazil}. \\
\textbf{Model-Specific Trends}: BLIP-2 exhibited a striking over-association, retrieving \textit{Bhutan} for nearly all positive traits among South Asians, often with extremely high percentages (80--98\%). This suggests a model collapse toward a singular national representation for a racial group. \\
\textbf{White Race Associations}: Unlike traditional Western dominance, \textit{Ukraine} surfaced as the most frequent country associated with positive traits for Whites across LACLIP, OpenCLIP, and BLIP-2, indicating a potential data or alignment shift toward Eastern Europe. \\
\textbf{Latino and Middle Eastern Groups}: Positive traits for Latino individuals were commonly linked to \textit{Chile} and \textit{Ecuador}, while Middle Eastern individuals were most associated with \textit{Egypt}, \textit{Jordan}, and \textit{Iran}.

\subsubsection{Negative Traits}
\textbf{Stereotypical Reinforcement}: Countries such as \textit{Philippines} (East Asian) and \textit{India} (South Asian) were frequently retrieved for traits like \textit{criminal} and \textit{lazy} across all models, exposing latent stereotypes embedded in VLMs. \\
\textbf{Black Race Negative Bias}: In ALIP, LACLIP, and OpenCLIP, \textit{South Africa} dominated the negative trait retrievals. However, BLIP-2 exhibited an even stronger and problematic bias, overwhelmingly associating all negative traits with the \textit{Democratic Republic of Congo (DRC)}, further highlighting a collapse into a singular representation. \\
\textbf{Middle Eastern and Latino Trends}: Middle Eastern countries such as \textit{Israel}, \textit{Iran}, and \textit{Jordan} were frequently retrieved for \textit{criminal} and \textit{dangerous}, pointing to the persistence of geopolitical stereotypes. Among Latinos, \textit{Paraguay}, \textit{Argentina}, and \textit{Ecuador} were common retrievals for negative traits.
\subsubsection{Model Comparison}
ALIP and LACLIP demonstrated relatively greater diversity and cultural nuance in their associations compared to OpenCLIP and BLIP-2. Although biases were still present, they were less extreme. OpenCLIP tended toward flattening associations but still retrieved stereotypical mappings. BLIP-2 showed the most severe overgeneralization for both positive and negative traits within racial categories, indicating critical fairness and robustness issues.

%Overall, the analysis reveals that current VLMs exhibit significant cultural biases in associating races and countries with personal traits. While ALIP and LACLIP show marginal improvements, all models display concerning patterns of stereotyping, overgeneralization, and representational collapse, particularly in underrepresented racial groups.

\section{Conclusion}
In this work we proposed a comprehensive framework for cultural evaluation of VLMs. Specifically we evaluated four models on three tasks such as race to country retrieval, physical traits to country retrieval and physical characteristic to country retrieval. Our results show that VLMs exhibhit stereotypical positive and negative biases towards just a few set of countries. These results reveal the lack of cultural understanding of VLMs across different races and nationalities.

{
    \small
    \bibliographystyle{ieeenat_fullname}
    \bibliography{main}
}

% WARNING: do not forget to delete the supplementary pages from your submission 
\clearpage
\setcounter{page}{1}
\maketitlesupplementary

\begin{figure*}[htbp]
    \centering

    \begin{subfigure}[t]{0.49\textwidth}
        \includegraphics[width=\linewidth]{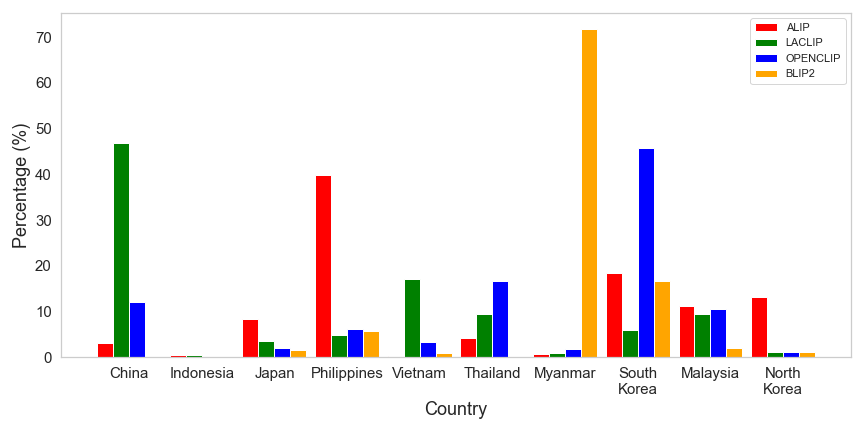}
        \caption{East Asian}
    \end{subfigure}
    \hfill
    \begin{subfigure}[t]{0.49\textwidth}
        \includegraphics[width=\linewidth]{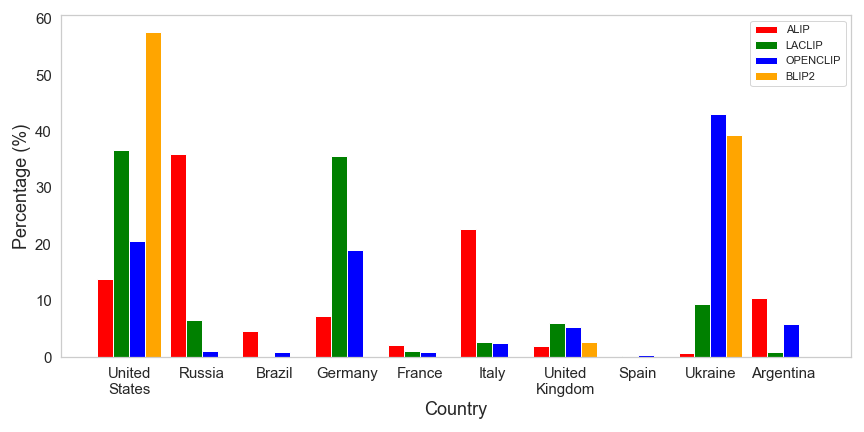}
        \caption{White}
    \end{subfigure}

    \vspace{0.5cm}

    \begin{subfigure}[t]{0.49\textwidth}
        \includegraphics[width=\linewidth]{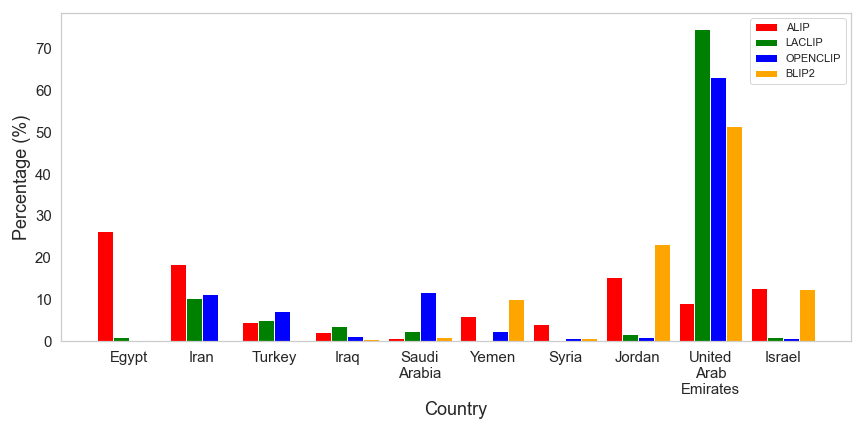}
        \caption{Middle Eastern}
    \end{subfigure}
    \hfill
    \begin{subfigure}[t]{0.49\textwidth}
        \includegraphics[width=\linewidth]{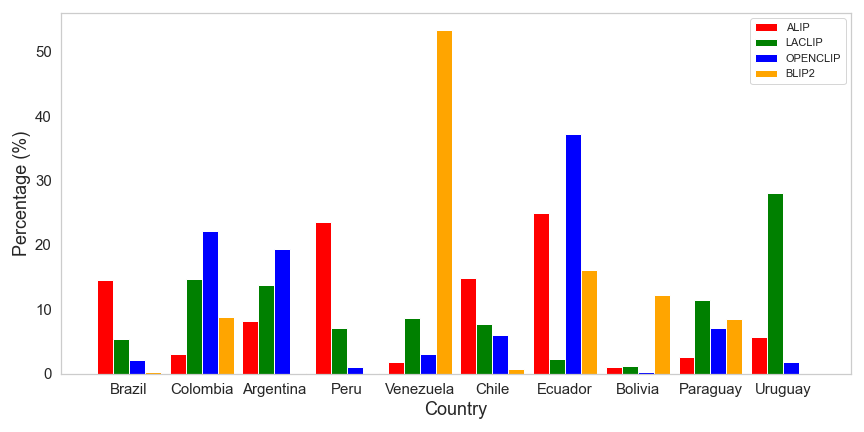}
        \caption{Latino}
    \end{subfigure}

    \vspace{0.5cm}

    \begin{subfigure}[t]{0.49\textwidth}
        \includegraphics[width=\linewidth]{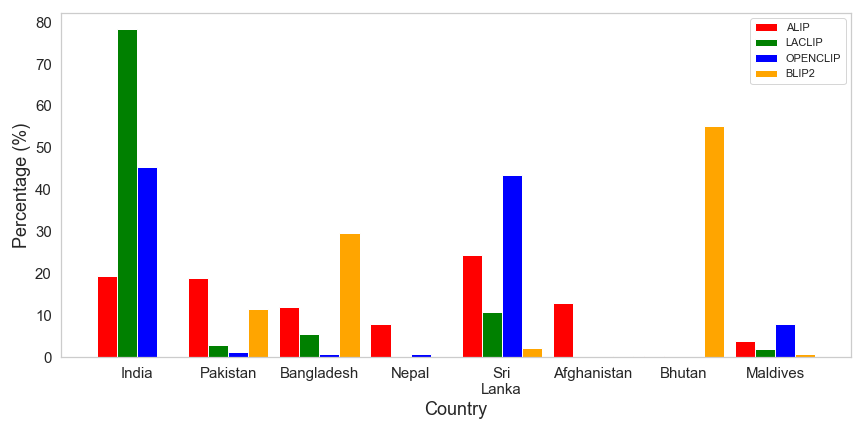}
        \caption{South Asian}
    \end{subfigure}
    \hfill
    \begin{subfigure}[t]{0.49\textwidth}
        \includegraphics[width=\linewidth]{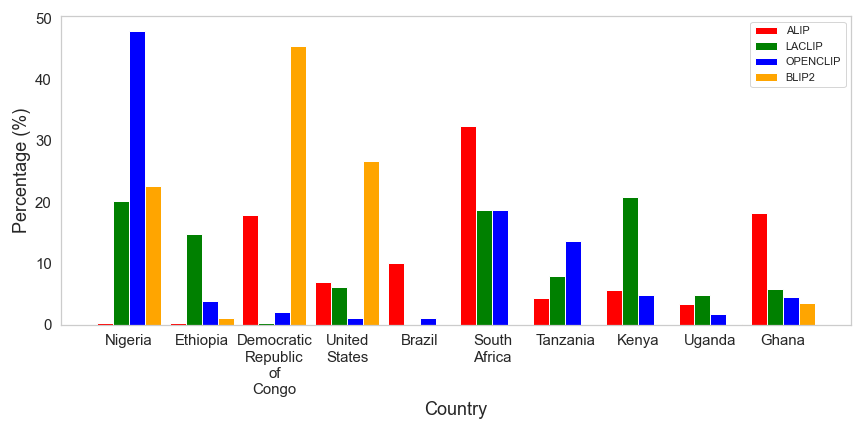}
        \caption{Black}
    \end{subfigure}

    \caption{\textbf{Performance comparison of different VLMs on various races and retrieved countries as a percentage of total images on Socialcounterfactuals dataset}}
    \label{fig:race_country_socialcounterfactuals}
\end{figure*}

\begin{table*}[t]
\centering
\tiny
\setlength{\tabcolsep}{1.8pt} % Reduces column separation
\renewcommand{\arraystretch}{1.3} % Adjust row spacing slightly if needed
\begin{tabular}{l|*{6}{c}|*{6}{c}}
\toprule
& \multicolumn{6}{c|}{ALIP} & \multicolumn{6}{c}{LACLIP} \\
\cline{2-13}
Trait & \makecell{East\\Asian} & \makecell{South\\Asian} & White & \makecell{Middle\\Eastern} & Latino & Black 
      & \makecell{East\\Asian} & \makecell{South\\Asian} & White & \makecell{Middle\\Eastern} & Latino & Black \\
\midrule
Smart & Philippines (49.46)  & Pakistan (44.3)  & Russia (58.96)  & Egypt (23.48)  & Chile (26.77)  & DRC (49.79)  & China (32.67)  & India (69.4)  & Ukraine (36.31)  & UAE (72.13)  & Peru (33.23)  & Nigeria (24.43) 
\\
Honest & Philippines (65.39)  & Afghanistan (35.19)  & Argentina (37.35)  & Jordan (38.6)  & Peru (41.05)  & South Africa (24.22)  & China (35.96)  & India (58.42)  & USA (34.44)  & UAE (83.93)  & Peru (28.78)  & Nigeria (23.77) 
\\
Successful & Philippines (31.68)  & India (22.35)  & Germany (43.79)  & Egypt (51.85)  & Peru (32.21)  & DRC (37.28)  & China (34.39)  & India (44.85)  & Ukraine (44.66)  & UAE (64.18)  & Peru (37.27)  & Nigeria (27.92) 
\\
Talented & Philippines (77.8)  & SriLanka (33.0)  & Italy (33.22)  & UAE (31.43)  & Brazil (39.44)  & DRC (30.73)  & China (53.61)  & India (44.63)  & USA (31.33)  & UAE (68.53)  & Peru (46.08)  & Nigeria (26.15) 
\\
Creative & Philippines (30.73)  & SriLanka (27.51)  & Russia (37.04)  & Israel (29.79)  & Chile (35.81)  & DRC (45.76)  & China (38.43)  & India (39.35)  & Ukraine (42.65)  & UAE (75.96)  & Peru (26.93)  & Uganda (26.25) 
\\
Hardworking & Philippines (56.22)  & Bangladesh (46.68)  & Russia (29.85)  & UAE (37.36)  & Uruguay (50.77)  & DRC (63.31)  & China (32.62)  & India (60.29)  & Ukraine (40.29)  & UAE (81.82)  & Uruguay (28.45)  & Nigeria (24.55) 
\\
Criminal & Philippines (29.51)  & SriLanka (40.22)  & France (35.81)  & Turkey (17.54)  & Bolivia (26.18)  & SouthAfrica (34.27)  & China (54.76)  & India (82.5)  & Germany (42.91)  & UAE (76.84)  & Peru (42.16)  & Ethiopia (27.99) 
\\
Lazy & Japan (38.93)  & India (27.62)  & Germany (53.47)  & Turkey (33.39)  & Uruguay (39.35)  & SouthAfrica (63.41)  & China (32.76)  & India (60.23)  & Ukraine (47.29)  & UAE (83.8)  & Paraguay (21.02)  & Ethiopia (29.56) 
\\
Dangerous & Philippines (29.93)  & SriLanka (40.55)  & France (32.47)  & Jordan (39.91)  & Brazil (57.53)  & Brazil (32.03)  & China (41.69)  & India (63.33)  & Germany (31.63)  & UAE (79.33)  & Uruguay (33.53)  & Ethiopia (25.68) 
\\
Poor & China (28.76)  & India (29.5)  & Germany (45.61)  & Jordan (52.71)  & Uruguay (32.02)  & SouthAfrica (25.47)  & China (44.13)  & India (68.46)  & Germany (43.17)  & UAE (79.4)  & Colombia (29.98)  & SouthAfrica (35.84) 
\\
Violent & Philippines (28.21)  & Maldives (34.29)  & Argentina (54.56)  & Egypt (25.55)  & Argentina (39.03)  & Brazil (44.42)  & China (53.13)  & India (79.21)  & Germany (35.5)  & UAE (59.12)  & Peru (42.7)  & Uganda (32.21) 
\\
Illiterate & Philippines (31.97)  & SriLanka (44.11)  & USA (51.66)  & UAE (37.82)  & Venezuela (40.95)  & DRC (36.53)  & China (48.25)  & India (69.21)  & USA (30.93)  & UAE (83.18)  & Uruguay (38.52)  & Ethiopia (30.0) 
\\
\bottomrule
\end{tabular}
\caption{\textbf{Comparison of personal traits across demographics for ALIP and LACLIP models on SocialCounterFactuals dataset.}}
\label{tab:trait_comparison_alip_laclip_socialcounterfactuals}
\end{table*}

\begin{table*}[t]
\centering
\tiny
\setlength{\tabcolsep}{1.8pt} % Reduces column separation
\renewcommand{\arraystretch}{1.3} % Adjust row spacing slightly if needed
\begin{tabular}{l|*{6}{c}|*{6}{c}}
\toprule
& \multicolumn{6}{c|}{OpenCLIP} & \multicolumn{6}{c}{BLIP-2} \\
\cline{2-13}
Trait & \makecell{East\\Asian} & \makecell{South\\Asian} & White & \makecell{Middle\\Eastern} & Latino & Black 
      & \makecell{East\\Asian} & \makecell{South\\Asian} & White & \makecell{Middle\\Eastern} & Latino & Black \\
\midrule
Smart & Vietnam (50.14)  & India (74.22)  & Germany (28.99)  & UAE (61.31)  & Ecuador (38.94)  & Nigeria (28.63)  & SouthKorea (63.8)  & Bangladesh (45.06)  & USA (43.57)  & Jordan (54.66)  & Colombia (75.48)  & DRC (54.67) 
\\
Honest & China (25.38)  & India (61.08)  & Ukraine (70.04)  & UAE (58.07)  & Ecuador (56.28)  & Nigeria (61.44)  & Philippines (31.27)  & SriLanka (56.5)  & Ukraine (49.03)  & Jordan (84.39)  & Colombia (82.25)  & Ghana (38.25) 
\\
Successful & Vietnam (52.53)  & India (64.79)  & Ukraine (37.12)  & UAE (53.7)  & Colombia (39.96)  & South Africa (35.01)  & Philippines (50.35)  & SriLanka (60.99)  & Ukraine (31.92)  & Jordan (78.52)  & Colombia (62.15)  & DRC (51.66) 
\\
Talented & Vietnam (36.44)  & India (74.93)  & USA (35.23)  & UAE (71.02)  & Venezuela (38.48)  & Nigeria (56.01)  & Japan (62.01)  & SriLanka (78.9)  & USA (42.48)  & Jordan (92.33)  & Colombia (77.79)  & Ghana (46.29) 
\\
Creative & Vietnam (40.4)  & India (69.2)  & Ukraine (50.37)  & UAE (79.23)  & Ecuador (28.86)  & Nigeria (50.39)  & Myanmar (27.34)  & SriLanka (76.51)  & Germany (58.39)  & Jordan (86.45)  & Colombia (70.5)  & DRC (61.45) 
\\
Hardworking & SouthKorea (39.74)  & India (78.11)  & Germany (58.52)  & UAE (41.11)  & Colombia (28.6)  & South Africa (55.61)  & Philippines (66.12)  & SriLanka (42.85)  & Ukraine (48.15)  & UAE (50.67)  & Colombia (65.45)  & DRC (76.48) 
\\
Criminal & China (41.41)  & India (77.84)  & Germany (39.28)  & UAE (55.25)  & Ecuador (33.22)  & SouthAfrica (62.44)  & Philippines (75.96)  & Bhutan (39.16)  & USA (63.67)  & UAE (90.01)  & Ecuador (42.65)  & DRC (58.6) 
\\
Lazy & North Korea (30.15)  & India (82.77)  & Ukraine (61.2)  & UAE (36.04)  & Ecuador (44.3)  & Nigeria (63.24)  & Myanmar (36.49)  & Bhutan (87.9)  & Ukraine (38.21)  & Jordan (82.48)  & Colombia (63.41)  & Ghana (37.45) 
\\
Dangerous & SouthKorea (41.82)  & India (70.89)  & USA (53.89)  & UAE (72.34)  & Venezuela (44.43)  & Nigeria (60.37)  & SouthKorea (42.36)  & Bhutan (50.87)  & Ukraine (53.13)  & Jordan (93.57)  & Colombia (84.51)  & Ghana (69.75) 
\\
Poor & China (29.75)  & India (49.78)  & Germany (50.84)  & UAE (59.56)  & Ecuador (28.52)  & SouthAfrica (40.53)  & SouthKorea (40.32)  & Bhutan (40.61)  & United Kingdom (34.25)  & Jordan (87.54)  & Colombia (66.44)  & DRC (80.49) 
\\
Violent & China (42.44)  & India (73.48)  & Germany (68.78)  & UAE (43.76)  & Peru (39.24)  & Ethiopia (36.09)  & Philippines (69.75)  & Bhutan (46.66)  & Ukraine (45.02)  & UAE (85.15)  & Colombia (94.97)  & DRC (48.42) 
\\
Illiterate & SouthKorea (24.74)  & SriLanka (37.97)  & USA (50.88)  & UAE (68.05)  & Chile (28.79)  & SouthAfrica (47.41)  & SouthKorea (75.48)  & SriLanka (46.76)  & Ukraine (74.19)  & SaudiArabia (46.74)  & Colombia (40.69)  & DRC (60.24) 
\\
\bottomrule
\end{tabular}
\caption{\textbf{Comparison of personal traits across demographics for OpenCLIP and BLIP-2 models on SocialCounterFactuals dataset.}}
\label{tab:trait_comparison_openclip_blip2_socialcounterfactuals}
\end{table*}

\begin{figure*}[t]
    \centering
    \includegraphics[width=\linewidth]{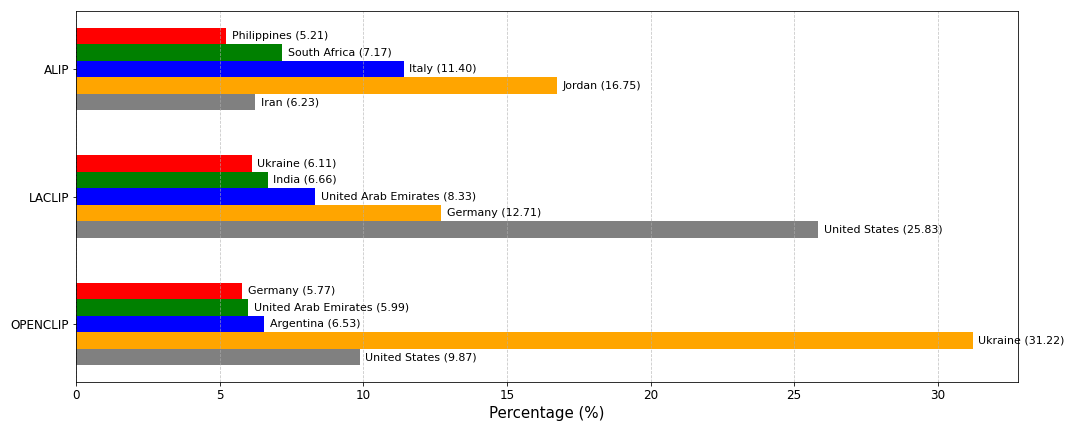}
    \caption{\textbf{Figure shows the top-5 countries retrieved for ALIP, LACLIP and OPENCLIP. The images retrieved are for skinny physical characteristics.}}
    \label{fig:skinny}
\end{figure*}

\begin{figure*}[t]
    \centering
    \includegraphics[width=\linewidth]{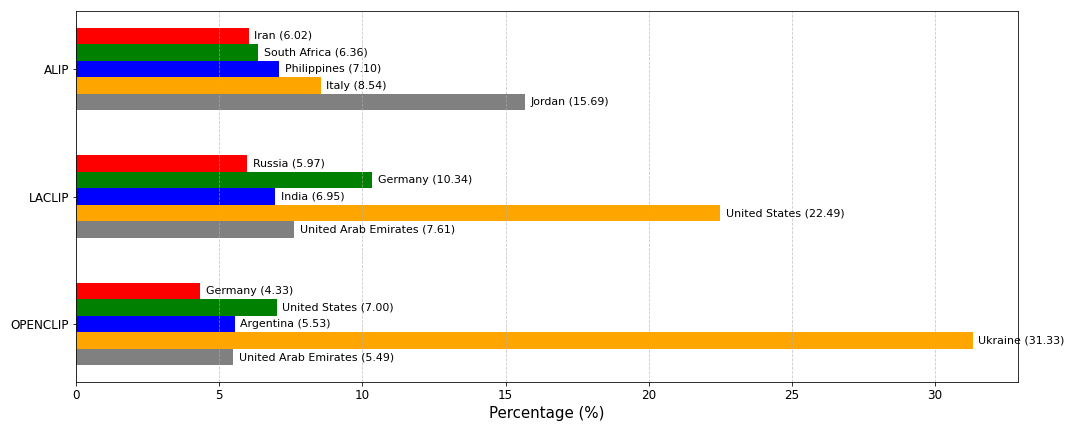}
    \caption{\textbf{Figure shows the top-5 countries retrieved for ALIP, LACLIP and OPENCLIP. The images retrieved are for young physical characteristics.}}
    \label{fig:young}
\end{figure*}

\begin{figure*}[t]
    \centering
    \includegraphics[width=\linewidth]{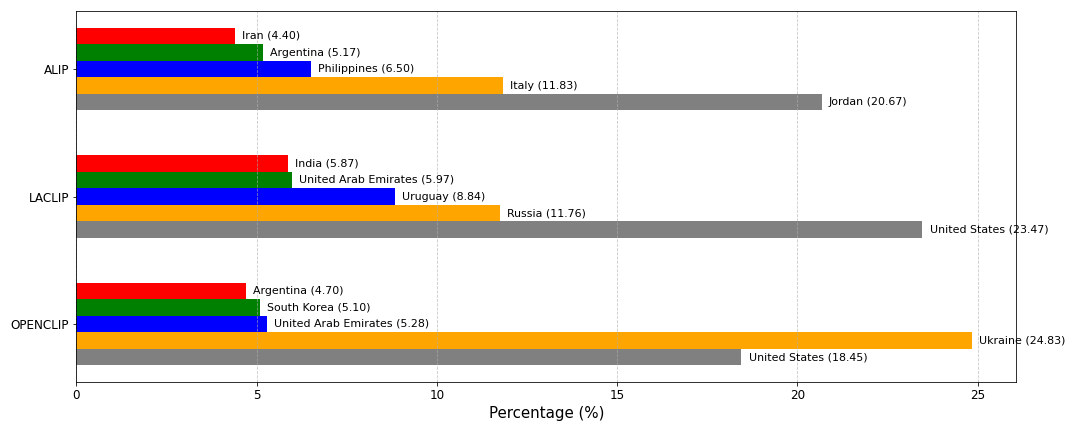}
    \caption{\textbf{Figure shows the top-5 countries retrieved for ALIP, LACLIP and OPENCLIP. The images retrieved are for obese physical characteristics.}}
    \label{fig:obese}
\end{figure*}

\begin{figure*}[t]
    \centering
    \includegraphics[width=\linewidth]{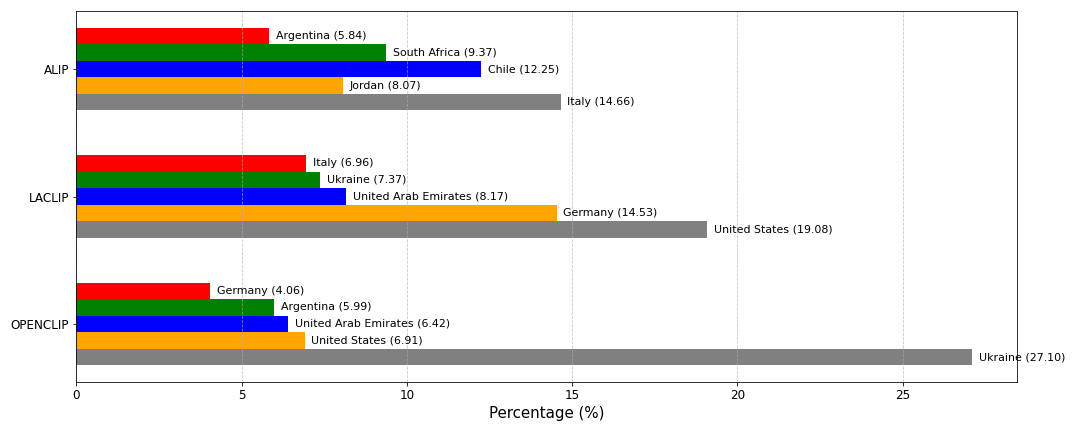}
    \caption{\textbf{Figure shows the top-5 countries retrieved for ALIP, LACLIP and OPENCLIP. The images retrieved are for skinny tattooed characteristics.}}
    \label{fig:tattooed}
\end{figure*}

\begin{figure*}[t]
    \centering
    \includegraphics[width=\linewidth]{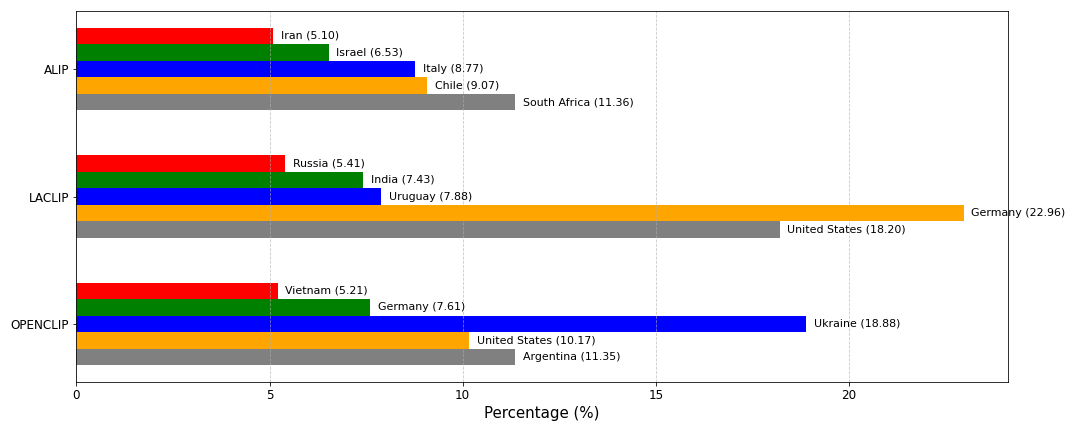}
    \caption{\textbf{Figure shows the top-5 countries retrieved for ALIP, LACLIP and OPENCLIP. The images retrieved are for old physical characteristics.}}
    \label{fig:old}
\end{figure*}

\section{Additional Results}
\subsection{VLMs encode persistent racial-regional biases across datasets}
To further test the robustness of these associations, we performed analysis on socialcounterfactuals dataset while retaining the same racial categories and neutral prompts. Figure ~\ref{fig:race_country_socialcounterfactuals} shows the results on this dataset. The results reaffirmed the presence of racialized regional biases, with BLIP-2 again showing sharp, isolated associations—favoring the United States for White individuals (57.7\%) and the Democratic Republic of Congo for Black individuals (45.6\%). ALIP distributed retrievals more evenly, though stereotypes persisted, linking Asian faces predominantly to the Philippines (39.8\%) and South Asian faces to India (19.3\%) and Sri Lanka (24.5\%). OpenCLIP demonstrated persistent preferences, notably associating Black faces with Nigeria (47.9\%) and Indian faces with India (45.4\%) and Sri Lanka (43.6\%). LACLIP maintained a comparatively balanced retrieval pattern, though regional clustering was still apparent—such as Asian faces around China and the Philippines, and Middle Eastern faces toward Iran and Turkey.

\subsection{Personal Traits to Country Retrieval.}
We also experiment personal traits to country retieval on SocialCounterFactuals dataset. Tables ~\ref{tab:trait_comparison_alip_laclip_socialcounterfactuals} and ~\ref{tab:trait_comparison_openclip_blip2_socialcounterfactuals} show the results for ALIP, LACLIP, OpenCLIP and BLIP-2 models. The findings reveal notable patterns of cultural attribution and bias across models. In ALIP, positive traits like \textit{smart}, \textit{honest}, \textit{successful}, and \textit{talented} for East Asians were most associated with the Philippines, suggesting a strong stereotypical mapping in the model's representation. South Asians were similarly associated with positive traits but through a different set of countries such as Pakistan, Afghanistan, and India, though with slightly lower confidence percentages compared to East Asians. White individuals were predominantly linked with European countries like Russia, Germany, and France for positive traits, reflecting a Western-centric bias in competence-related attributes. For Middle Eastern and Latino groups, associations were more mixed, but traits like \textit{successful} and \textit{hardworking} were often linked to countries like UAE, Peru, and Uruguay, indicating a modest positive bias.

Negative traits, however, showed clearer evidence of problematic stereotyping. In ALIP, \textit{criminal}, \textit{lazy}, and \textit{dangerous} traits were more frequently associated with South Asian and Black images. For instance, Black individuals were heavily associated with the Democratic Republic of Congo (DRC) across multiple negative traits such as \textit{dangerous} and \textit{criminal}. Latino groups showed a tendency to be matched with lower economic status attributes such as \textit{poor} or \textit{illiterate}, with countries like Uruguay and Venezuela frequently appearing.

LACLIP displayed both continuities and deviations from ALIP. Although it generally reduced the extent of extreme associations, it still presented patterns of cultural bias. Notably, LACLIP showed higher positive associations for Middle Eastern individuals compared to ALIP, linking them with traits like \textit{hardworking} and \textit{successful} with countries such as UAE at very high confidence scores ($>80\%$). Nonetheless, Black individuals continued to be strongly associated with Nigeria and Ethiopia for negative traits, suggesting persistent racial biases that are resistant across architectures.

OpenCLIP presented a somewhat different picture. The model was relatively more consistent in associating positive traits with South Asians, predominantly India, across almost all positive attributes. Confidence scores were notably high, often exceeding $70\%$, for positive traits such as \textit{smart}, \textit{successful}, and \textit{hardworking}. However, for negative traits, Black and Middle Eastern groups were still more heavily penalized, with countries like Nigeria, South Africa, and UAE frequently appearing for \textit{criminal} and \textit{dangerous} traits. OpenCLIP’s results suggest that while it is capable of strong positive cultural attributions, it remains vulnerable to negative racial profiling.

BLIP-2, the most recent and arguably the most powerful model among the four, exhibited the most polarized results. On positive traits, Latino individuals were overwhelmingly associated with Colombia across traits like \textit{smart}, \textit{successful}, \textit{creative}, and \textit{hardworking}, with extremely high matching percentages often above $75\%$. This points towards a more homogenized cultural mapping in BLIP-2, where a single country dominates the representation of positive traits for a racial group. For Black individuals, the Democratic Republic of Congo continued to be linked with both positive (\textit{hardworking}, \textit{creative}) and negative (\textit{dangerous}, \textit{criminal}) traits. Interestingly, in contrast to other models, BLIP-2 associated Black individuals with positive traits at a higher rate than earlier models but still maintained stereotypes for negative traits, indicating a partial but incomplete mitigation of bias.

Across all models, White individuals were frequently linked with Western nations like Germany, Ukraine, and the USA for both positive and, to a lesser extent, neutral or mildly negative traits. There were comparatively fewer strong associations with overtly negative traits like \textit{criminal} or \textit{lazy}, reflecting an underlying racial advantage embedded in the visual-textual representation learned by these models.

Although models like LACLIP and BLIP-2 show improvements in associating positive traits across diverse groups, a persistent trend remains where negative attributes are disproportionately associated with South Asian, Middle Eastern, and Black racial representations. Moreover, the dominance of specific countries such as the Philippines, India, Colombia, and the Democratic Republic of Congo across several traits highlights the risk of cultural oversimplification and stereotyping in vision-language models.
\subsection{Analysis of country retrievals across body types}
We study the cultural association of VLMs to physical characteristics of persons. Specifically we evaluate body types such as skinny, young, obese, tattooed and old. We use the SocialCounterFactuals dataset for this setup where the input is an image of a person and match it with prompts from all the countries mentioned in table ~\ref{tab:race_countries}. For each model, we then select the top-5 countries retrieved as a measure of the highest percentage. The results are shown in the figures ~\ref{fig:skinny}, ~\ref{fig:young}, ~\ref{fig:obese}, ~\ref{fig:tattooed} and ~\ref{fig:old} respectively.

The experimental results reveal distinct patterns in how different vision-language models (ALIP, LACLIP, and OPENCLIP) associate body types with specific countries, highlighting potential cultural biases embedded in these systems.
\subsubsection{Country association patterns}
Our analysis across five body types (skinny, young, obese, tattooed, and old) demonstrates consistent bias patterns within each model family, but with notable variations between models. OPENCLIP shows a strong tendency to associate Ukraine with multiple body types, particularly in the skinny, obese, and tattooed categories (31.22\%, 31.33\%, and 24.83\% respectively). This over-representation suggests potential dataset biases or embedding space distortions specific to this model architecture.
The United States appears prominently across all models, with particularly high representation in LACLIP results (25.83\% for skinny, 22.49\% for young, and 25.47\% for obese). This Western-centric bias likely reflects training data composition that oversamples American or Western imagery, creating an implicit association between various body types and American identity.
\subsubsection{Body type specific associations}
For the skinny body type (Image 1), Jordan shows significant representation in ALIP (16.75\%), while LACLIP heavily favors the United States (25.83\%). This divergence suggests fundamentally different conceptual associations between thinness and national identity across model architectures.
Young body types (Image 2) show different top countries across models, with Jordan dominating in ALIP (15.69\%), the United States in LACLIP (22.49\%), and Ukraine in OPENCLIP (31.33\%). This heterogeneity in results indicates unstable representations of youth across cultural contexts.
For obese body types (Image 3), Jordan (20.67\%) and the United States (25.47\%) dominate ALIP and LACLIP respectively, while OPENCLIP presents a more balanced distribution between Ukraine (24.83\%) and the United States (18.45\%). These associations may reflect stereotypical Western representations of obesity prevalence.
Tattooed bodies (Image 4) show strong associations with Italy in ALIP (14.66\%), while LACLIP associates them primarily with the United States (19.08\%) and Germany (14.53\%). OPENCLIP strongly associates tattoos with Ukraine (27.10\%). These variations likely reflect different cultural narratives about body modification across training datasets.
Elderly representation (Image 5) shows associations with South Africa (11.36\%) in ALIP, while Germany dominates LACLIP results (22.96\%) and Ukraine leads in OPENCLIP (18.88\%). These variations suggest different cultural framings of aging across the models' training data.

\end{document}